\documentclass{article}
\usepackage{INTERSPEECH2020}
\usepackage{amsmath,graphicx}
\usepackage{graphics}
\usepackage{color}
\usepackage{multirow}
\usepackage{url}
\usepackage{balance}
\usepackage{enumitem}

\newcommand{\omitthis}[1]{}
\ninept

\def\diag{\operatorname{diag}}
\title{SEQUENTIAL MULTI-FRAME NEURAL BEAMFORMING\\FOR SPEECH SEPARATION AND ENHANCEMENT}
\name{Zhong-Qiu Wang$^{1,2,*}$, Hakan Erdogan$^1$, Scott Wisdom$^1$, Kevin Wilson$^1$, \\Desh Raj$^3$, Shinji Watanabe$^3$, Zhuo Chen$^4$, John R.\ Hershey$^1$  
\thanks{*Work done during an internship at Google.}}
\address{$^1$Google Research, Cambridge, MA, $^2$MERL, Cambridge, MA\\$^3$Johns Hopkins University, Baltimore, MD, $^4$Microsoft, Seattle, WA}
\email{wang.zhongqiu41@gmail.com, \{hakanerdogan, scottwisdom, kwwilson, johnhershey\}@google.com, draj@cs.jhu.edu, shinjiw@ieee.org, zhuc@microsoft.com}

\begin{document}
%
\maketitle
\begin{abstract}
This work introduces sequential neural beamforming, which alternates between neural network based spectral separation and beamforming based spatial separation. Our neural networks for separation use an advanced convolutional architecture trained with a novel stabilized signal-to-noise ratio loss function. For beamforming, we explore multiple ways of computing time-varying covariance matrices, including factorizing the spatial covariance into a time-varying amplitude component and a time-invariant spatial component, as well as using block-based techniques. In addition, we introduce a multi-frame beamforming method which improves the results significantly by adding contextual frames to the beamforming formulations. We extensively evaluate and analyze the effects of window size, block size, and multi-frame context size for these methods. Our best method utilizes a sequence of three neural separation and multi-frame time-invariant spatial beamforming stages, and demonstrates an average improvement of 2.75 dB in scale-invariant signal-to-noise ratio and 14.2\% absolute reduction in a comparative speech recognition metric across four challenging reverberant speech enhancement and separation tasks. We also use our three-speaker separation model to separate real recordings in the LibriCSS evaluation set into non-overlapping tracks, and achieve a better word error rate as compared to a baseline mask based beamformer.
\end{abstract}
\section{Introduction}
\label{sec:intro}


Audio source separation has many applications, for example as a front end for robust automatic speech recognition (ASR) and to improve voice quality for telephony. 
Leveraging multiple microphones has great potential to improve separation, since the spatial relationship among microphones provides 
complementary information to spectral cues exploited by monaural 
approaches.
Multi-microphone processing can also improve the suppression of reverberation and diffuse background noise.


Recently, a new paradigm has emerged as a promising alternative to conventional beamforming approaches: neural beamforming, where the key advance is to utilize the non-linear modeling power of deep neural networks (DNN) to identify time-frequency (T-F) units dominated by each source for spatial covariance matrix computation \cite{heymann2016neural, erdogan2016improved}. Unlike traditional approaches, neural beamforming methods have the potential to learn and adapt from massive training data, which improves their robustness to unknown positions and orientations of microphones and sources, types of acoustic sources, and room geometry. An initial success of neural beamforming approaches was improving time-invariant beamforming using T-F domain mask prediction, where predicted masks were used to obtain time-invariant spatial covariance matrices for all sources. This has proven useful in ASR tasks such as CHiME-3/4 \cite{Barker2017}. Recent studies considered online or low-latency beamforming \cite{boeddeker2018exploring, yoshioka2019low} and time-varying beamforming \cite{kubo2019mask} for improved performance in certain scenarios. In addition, spatial features such as inter-channel phase differences (IPD) \cite{yoshioka2018multi}, cosine and sine IPDs \cite{wang2018multi} and target direction compensated IPDs \cite{wang2018combining}, which can encode spatial information, are utilized as additional network input to improve the mask estimation in masking-based beamforming. Other cues, such as visual information \cite{ephrat2018looking}, location information \cite{chen2018multi} and speaker embeddings \cite{delcroix2018single,wang2018voicefilter}, can also be used as additional inputs to improve target extraction and source separation in both single- and multi-microphone setups. 






This paper explores alternating between spectral estimation using DNN-based masking and spatial separation using linear beamforming with a multichannel Wiener filter (MCWF), performing up to three applications of the neural separation network: separate, beamform, separate, beamform, and separate. By doing so, linear beamforming is effectively driven by DNN-based masking.  This is inspired by the single-channel sequential network of \cite{kavalerov2019universal}, which we use as a baseline, and by the findings that better beamforming results can be used as extra features to improve spectral masking and vice versa. This sequential approach is related to iterative neural beamforming with postfiltering \cite{zhang2017speech}, which uses the same DNN repeatedly with only the beamformed signal as input for speech enhancement; in contrast, we train a different neural network for each stage, concatenating the mixture signal with the beamformed signals, applied to both speech enhancement and separation.
For beamforming, we consider both time-invariant and time-varying ways of calculating covariance matrices to improve spatial separation. We also explore the effectiveness by incorporating multi-frame context during beamforming. Evaluation results on four challenging sound separation tasks demonstrate the effectiveness of the proposed algorithms. Also, our best three-speaker separation model achieves significantly improved word error rate on the LibriCSS dataset compared to a neural mask beamforming baseline separation system.  






\section{Contributions}
The model we introduce has similarities with earlier mask-based beamforming models but has the following novel aspects which end up improving the performance significantly.
\begin{itemize}[leftmargin=4mm]
    \item We perform multiple repetitions of mask-based beamforming where each sequential application of a neural network has its own parameters optimized separately.
    \item We use different window and hop sizes for the mask-prediction network and the beamformer, in contrast to previous works where the same STFT parameters were used for both. Our networks predict time-domain waveforms, and we take another independent STFT to perform beamforming. We achieve best results using a smaller window size in the mask network and a larger one for the beamformer.
    \item We use a state-of-the-art TDCN++ network \cite{kavalerov2019universal} for mask-prediction as well as mixture consistency projection \cite{Wisdom2018} to improve separation performance.
    \item We use a stabilized SNR loss function for training the mask-prediction neural network.
    \item Our best performing beamformer is a time-invariant multi-frame multichannel Wiener filter that improves the results significantly compared to previously-used single-frame beamformers.
    \item Our model is completely independent of microphone and room geometry, and we show that our three-speaker separation model works well on a mismatched real data set with different number of microphones and an unseen room geometry.
\end{itemize}

\section{Methods}
\label{sec:method}


Assume an $M$-channel time-domain signal consisting of $S$ sources, $\mathbf{y}[n]=\sum_{s=1}^{S}\mathbf{x}^{(s)}[n]$, recorded in a reverberant environment. The short-time Fourier transform (STFT) of this multichannel signal can be written as $\mathbf{Y}_{t,f}=\sum_{s=1}^{S}\mathbf{X}^{(s)}_{t,f}$,
where ${\mathbf{Y}}_{t,f}$ and ${\mathbf{X}}^{(s)}_{t,f} \in \mathbb{C}^M$ respectively represent the mixture and the reverberant image of source $s$ at time $t$ and frequency $f$. 
Our study proposes multiple algorithms to recover the constituent reverberant sources $X_\mathrm{ref}^{(s)}$ from a reverberant mixture $Y_\mathrm{ref}$ received by a reference microphone, with or without leveraging spatial information contained in $\mathbf{Y}$.
We assume offline processing and non-moving sources throughout each utterance.

\begin{figure}  
	\centering  
	\includegraphics[width=\linewidth]{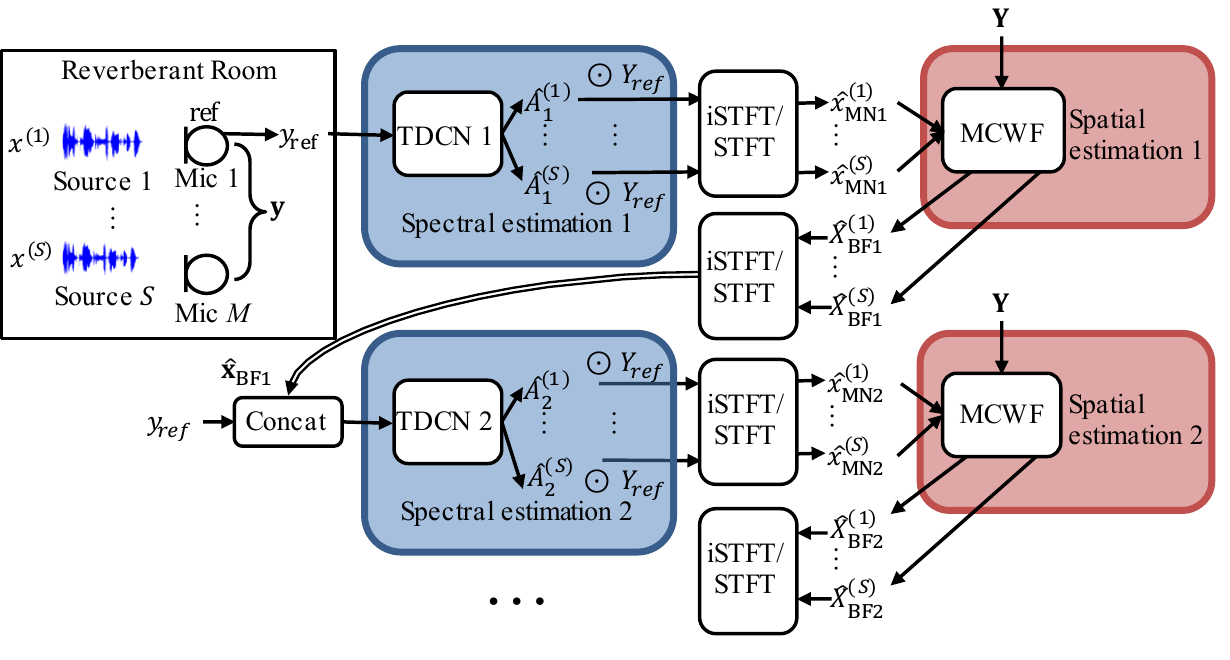}
 	\caption{System overview.}
	\label{fig:system}
    \vspace{-10pt}
\end{figure}

Figure \ref{fig:system} illustrates our proposed system. Each spectral masking stage uses an improved time-domain dilated convolutional neural network (TDCN++) \cite{kavalerov2019universal}.
The first stage performs single-channel processing to estimate each source via T-F masking. The estimated sources are then used to compute statistics for time-invariant or time-varying beamforming. The next masking stage combines spectral and spatial information by taking in the mixture and beamformed results for post-filtering. This sequence is then repeated several times.

As shown in Figure \ref{fig:system}, we train through multiple iSTFT/STFT projection layers. These layers are helpful as they can effectively address the well-known phase inconsistency problem, a common issue of magnitude-domain masking \cite{Wisdom2018, Wang2019}.
In addition, our masking networks operate at a typical 32 ms window size, but our system can use a larger window size for beamforming. This way, beamforming can be performed at a higher frequency resolution and produce finer separation. The iSTFT/STFT pairs are necessary here to change the window size back and forth during sequential processing.
This strategy dramatically improves time-invariant MCWF in our experiments. 

\subsection{Spectral mask estimation for sound separation}
For monaural speech enhancement and speaker separation, we use TDCN++ based T-F masking (see \cite{Wang2018review} for an overview) to produce source estimates $\hat{X}^{(s)}_{\mathrm{MN}i} = \hat{A}^{(s)}_{i}\odot Y_{\mathrm{ref}}$, where $\odot$ denotes element-wise multiplication and $\hat{A}^{(s)}_{i}$ is the mask estimate produced by the $i$th TDCN++. Note that $i\in\{1,2,...,I\}$, meaning that there are $I$ (set to $3$ in this study) stages in the sequence. 
The loss function at each stage maximizes a stabilized SNR in the time domain
\begin{equation}
\mathcal{L}_{i} = 
\min_{\pi \in \Pi} 
\sum_{s=1}^{S} 
-\textrm{SNR}_{\tau,\epsilon}\Big(
    \textrm{iSTFT}(
        \hat{X}^{(\pi(s))}_{\mathrm{MN}i}
    ),
    {x}_\mathrm{ref}^{(s)}
\Big),
\label{sc-loss}
\end{equation}
where
\begin{equation}
    \textrm{SNR}_{\tau,\epsilon}(\hat{\bf x}, {\bf x})
    =
    10\log_{10}\left(
    \frac
        {\|{\bf x}\|^2}
        {\|{\bf x}-\hat{\bf x}\|^2 + \tau \|\bf x\|^2 + \epsilon}
    \right),
\end{equation}
$\Pi$ is the set of all permutations over $S$ sources, and $\pi(s)$ is the permuted source index $s$ under permutation $\pi$. The parameter $\tau$ 
acts as a soft threshold limiting the maximum SNR that can dampen the effect on the total loss from examples that are already well-separated. We use permutation invariance 
for speech separation, but not for speech enhancement. 
For $i > 1$, the network input is the concatenation of the mixture magnitude STFT features with those of all beamformed source estimates, $\hat{X}^{(s)}_{\mathrm{BF}i-1}$.
\subsection{Multi-frame multichannel Wiener filter}
Inspired by the success of convolutional beamformers \cite{nakatani2019unified}, we introduce multi-frame MCWF beamforming and investigate various context sizes.
The rationale is that, by stacking multiple frames, the beamformer can have more contextual information and degrees of freedom for better noise suppression.

We first define a context-expanded observed signal
$\Bar{\mathbf{Y}}_{t,f} = \left[\mathbf{Y}^T_{t-a,f}, \ldots, \mathbf{Y}^T_{t,f}, \ldots, \mathbf{Y}^T_{t+b,f}\right]^T \in \mathbb{C}^{cM}$
which is a flattened complex vector including multiple frames around a T-F unit, where $a$ is the left 
and $b$
the right context size in frames, and $c=a+b+1$. 
We treat all the contextual T-F units as if they are additional microphones in the subsequent beamforming formulations.
In each stage $i$, estimated sources $\hat{X}^{(s)}_{\mathrm{MN}i}$ from the TDCN++ are used to compute the spatial covariance of each source for a time-invariant MCWF (TI-MCWF)
\begin{equation}
\hat{\mathbf{w}}^{(s)}_{i,f} = (\hat{\mathbf{\Phi}}^{(y)}_{f})^{-1} \hat{\mathbf\Phi}^{(s)}_{i,f} \mathbf{u}_\mathrm{ref},
  \label{ti-mcwf}
\end{equation}
where $\mathbf{u}_\mathrm{ref}$ is a one-hot vector with the coefficient corresponding to the reference microphone at the center frame set to one, the multi-frame mixture covariance matrix is estimated as
\begin{equation}
{\hat{\mathbf\Phi}}^{(y)}_f=\dfrac{1}{T} \sum_{t=1}^{T} \Bar{\mathbf{Y}}_{t,f}\Bar{\mathbf{Y}}_{t,f}^{H},
\end{equation}
and $\hat{\mathbf\Phi}^{(s)}_{i,f}$ is the source covariance matrix computed as
\begin{equation}
\hat{\mathbf\Phi}^{(s)}_{i,f} = \dfrac{1}{T} \sum_{t=1}^{T} \hat{A}^{(s)}_{i,t,f}\Bar{\mathbf{Y}}_{t,f}\Bar{\mathbf{Y}}_{t,f}^{H},
  \label{ti-c-cov}
\end{equation}
\begin{equation}
\hat{A}^{(s)}_{i,t,f}=\dfrac{|\hat{X}^{(s)}_{\mathrm{MN}i,t,f}|^2}{\sum_{s'=1}^{S} |\hat{X}^{(s')}_{\mathrm{MN}i,t,f}|^2}.
  \label{mask}
\end{equation}
This approach follows recent developments in neural beamforming \cite{Barker2017,Heymann2015,erdogan2016improved} and straightforwardly applies them to a multi-frame setup.
The idea is to use T-F units dominated by source $s$ to compute its covariance matrix for beamforming. Here the Wiener-like mask $\hat{A}^{(s)}$, which can be derived based on a different window size, is recomputed in an alternate STFT domain from the reconstructed time-domain signal for source $s$ from the masking network. For convenience, the mask is considered the same across microphones, which is a 
reasonable 
approximation for compact arrays in far-field conditions. The beamforming result for source $s$ in stage $i$ is computed as
\begin{equation}
\hat{X}^{(s)}_{\mathrm{BF}i,t,f} = (\hat{\mathbf{w}}^{(s)}_{i,f})^{H}\Bar{\mathbf{Y}}_{t,f}.
  \label{ti-mcwf-apply}
\end{equation}

We also experimented with MVDR and MPDR beamformers \cite{souden2009optimal, ehrenberg2010sensitivity}, but they did not perform as well as MCWF in terms of SI-SNR. This paper hence only reports results with MCWF.

\subsection{Time-varying beamforming for spatial estimation}
A TI-MCWF has limited power for separation, as it is only a linear time-invariant filter per-frequency. To obtain time-varying behavior, we experiment with a block-based approach, where we calculate TI-MCWF beamformers in half-overlapping blocks of frames with some windowing. We use windowed signals to calculate spatial covariance matrices and perform overlap-add for post-windowed beamformed signals.  The Vorbis window \cite{valin2016high} is used for this processing.

\omitthis{
Similar to recursive averaging, one straightforward way to improve the separation capability is to estimate the covariances $\Bar{\mathbf\Phi}^{(y)}$ and $\Bar{\hat{\mathbf\Phi}}^{(s)}$ within a sliding window, i.e.
\begin{equation}
\mathbf\Phi^{(y)}_t=\dfrac{1}{2\Delta+1} \sum_{t'=t-\Delta}^{t+\Delta} \mathbf{Y}_{t'}\mathbf{Y}_{t'}^{H}
  \label{phi-y-sliding}
\end{equation}
\begin{equation}
\hat{\mathbf\Phi}^{(s)}_t = \dfrac{1}{2\Delta+1} \sum_{t'=t-\Delta}^{t+\Delta} \hat{A}^{(s)}(t',f)\mathbf{Y}(t',f)\mathbf{Y}(t',f)^{H}
  \label{phi-c-sliding}
\end{equation}
where $\Delta$ is half the window size in frames and $\hat{M}^{(c)}$ from (\ref{mask}).
}

A frame-level way of computing a time-varying covariance matrix for each source is to factorize it as a product of a time-varying power spectral density (PSD) and a time-invariant coherence matrix \cite{Duong2010, Higuchi2016, Shimada2019}. The rationale is that for a non-moving source, its coherence matrix is time-invariant assuming that the beamforming STFT window is long enough to capture most of the reverberation. Unlike conventional methods, which typically use maximum likelihood estimation or non-negative matrix factorization to estimate the PSD and spatial coherence \cite{Duong2010, Shimada2019}, the proposed algorithm leverages estimated source signals produced by neural networks to compute these statistics. Mathematically,
\begin{equation}
\hat{\mathbf\Phi}^{(s)}_{i,t,f} 
= 
|\hat{X}^{(s)}_{\mathrm{MN}i,t,f}|^2 
\hat{\mathbf\Psi}^{(s)}_{i,t,f} / \hat{D}_{i,t,f},
  \label{tv-c-cov}
\end{equation}
where
$|\hat{X}^{(s)}_{\mathrm{MN}i,t,f}|^2$
is the PSD estimate,
$\hat{\mathbf\Psi}^{(s)}_{i,t,f}$
can be either
$\hat{\mathbf\Phi}^{(s)}_{i,f}$
computed over all the frames in an utterance as in (\ref{ti-c-cov}) for a time-invariant covariance matrix, or it could be a block-based one calculated over the 
frames in a block.
$\hat{D}_{i,t,f} = \hat{d}_{i,t,f} \hat{d}_{i,t,f}^{T}$ 
with
$\hat{d}_{i,t,f} = \diag(\hat{\mathbf\Psi}^{(s)}_{i,t,f})^{1/2}$ 
normalizes the spatial component to have a unit diagonal. In far-field conditions where level differences are negligible,
$\hat{D}_{i,t,f} \approx (\hat{\Psi}^{(s)}_{i,t,f})_{m,m} \mathbf{1} \mathbf{1}^T$
for a microphone index $m$.
A time-varying factorized (TVF) MCWF is computed as
\begin{equation}
\hat{\mathbf{w}}^{(s)}_{i,t,f} = (\hat{\mathbf\Phi}^{(y)}_{i,t,f})^{-1} \hat{\mathbf\Phi}^{(s)}_{i,t,f} \mathbf{u}_\mathrm{ref},
  \label{tv-mcwf}
\end{equation}
where $\hat{\mathbf\Phi}^{(y)}_{i,t,f}=\sum_{s'=1}^{S}\hat{\mathbf\Phi}^{(s')}_{i,t,f}$, and the beamformed result is
\begin{equation}
\hat{X}_{\mathrm{BF}i,t,f}^{(s)} = (\hat{\mathbf{w}}^{(s)}_{i,t,f})^{H}\Bar{\mathbf{Y}}_{t,f}.
  \label{tv-mcwf-apply}
\end{equation}

\section{Data and models}
\label{sec:setup}
\subsection{Datasets}
We use room impulse responses (RIRs) generated by an image-method room simulator with frequency-dependent wall filters. For each example, the RIR is created by sampling random locations for a cube-shaped microphone array and all sources within a room defined using a random size: width from 3 to 7 m, length from 4 to 8 m, and height between 2.13 and 3.05 m. 
The sides of the cube-shaped array was 20 cm long.
During RIR generation, all source "image" locations are randomly perturbed by up to 8 cm in each direction to avoid the ``sweeping echo'' effect \cite{de2015modeling}.
We generate 140,000 training, 20,000 validation and 20,000 test rooms which are used to generate train, validation and test data.
Clean speech is from Libri-Light \cite{kahn2020librilight} and LibriTTS \cite{zen2019libritts}, and non-speech sounds are from \url{freesound.org}.
We filtered out artificial sounds (such as synthesizer noises) based on user-annotated tags and used a sound classification network trained on AudioSet \cite{AudioSet} to avoid clips likely containing speech.
During training, sources are reverberated and mixed on the fly, 
and the validation and test sets consist of about 10 hours of mixture data each. Recipes for these datasets will be publicly released in the near future. We validate the proposed algorithms on 1, 2, 4, and 8-microphone setups. 

Using this source data, the proposed models are evaluated on both speech separation and speech recognition. For speech separation evaluation, we construct three tasks: two-speaker separation, three-speaker separation and speech enhancement. For the speech enhancement task, a speech source is mixed with three directional noise sources, and the goal is to separate the speech from the noises. For each task, a random speech clip from clean source data is selected, and then each of the other sources is scaled to an SNR randomly drawn from $\mathcal{N}(0,7)$ dB with respect to the speech clip.
To better compare with previous arts, we used an additional two-speaker separation evaluation dataset introduced in \cite{wang2018multi}, which is a multichannel reverberated version of WSJ0-2mix database simulated using a room simulator with random room configurations and microphone positions. 

Besides these separation tasks, we evaluate our three-speaker separation model in terms of its ASR performance on the LibriCSS dataset which is a real meeting-like overlapping speech dataset \cite{chen2020continuous}. This dataset has been collected by playing LibriSpeech utterances from loudspeakers and recording them in a room. Each loudspeaker takes a role of a single speaker and reads only utterances from that speaker \cite{chen2020continuous}.

\subsection{Why use simulated data?}
Our algorithms require multichannel data and we would like to make sure that we see a huge number of possible mixing configurations during training. Thus, we use simulation to generate training data with a multitude of possible source locations and microphone positions in random rooms. We have to use this type of simulated data since existing ``real'' multichannel recordings and room impulse response databases are nowhere close to the size that's needed to train a good separation model that generalizes to unseen conditions at test time. For example, the ACE database \cite{eaton2015ace} and BUT ReverbDB \cite{Szoke_2019} provide real room impulse responses, but they are limited in the number of rooms, possible source locations, and range of microphone geometries: ACE has 7 rooms, 1 source location per room, and one geometry each for 2, 3, 4, 5, 8, and 32 mics; BUT ReverbDB has 8 rooms, 2-11 source positions per room with a spherical 8-mic array in addition to 23 single microphones with ad-hoc placements.  Thus, these real RIR databases are not extensive enough for training generalizable multichannel separation models that can handle arbitrary numbers of sources in arbitrary rooms. For evaluation of separation systems, simulated test sets provide ground-truth source references that can be used to measure performance in terms of SI-SNR. Due to scarcity of common simulated multichannel evaluation sets, we generate our own development and evaluation sets. We plan to release our simulated databases to allow wide use of the academic community and serve as a benchmark for multichannel separation tasks. Real multichannel datasets with transcription but without ground-truth source reference signals are available. To provide evaluation on real data in this paper, we evaluate our best model with ASR word error rates on the LibriCSS dataset, which has real acoustic mixing and known transcriptions. 

\begin{figure*}[htb]
\centering
\begin{minipage}[b]{0.42\linewidth}
  \centering
  \centerline{\includegraphics[width=\linewidth]
  {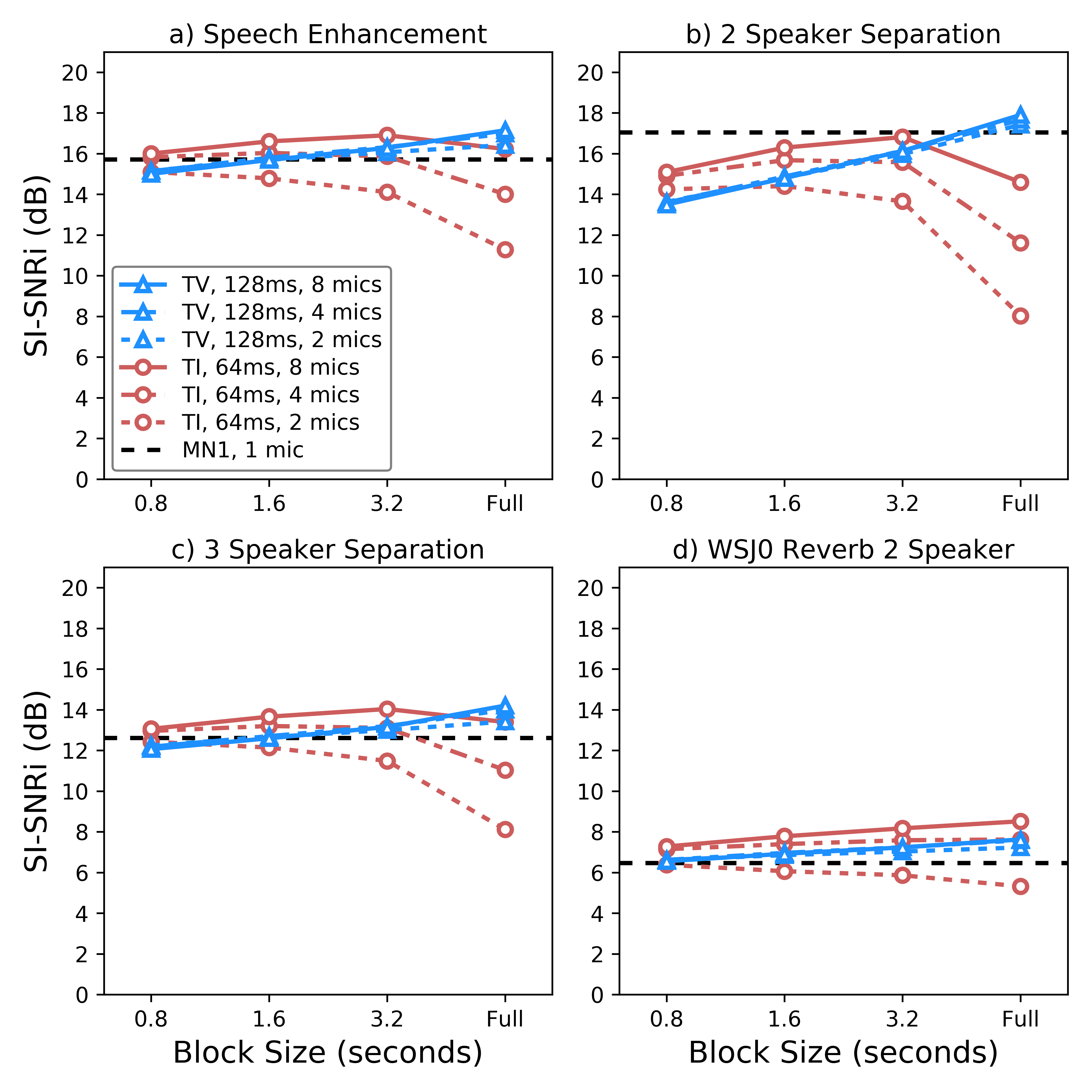}}
\vspace{-10pt}
\caption{Single-frame beamforming {versus} block size.}
\label{fig:bf_single}
\end{minipage}
\hspace{0.025\linewidth}
\begin{minipage}[b]{0.42\linewidth}
  \centering
  \centerline{\includegraphics[width=\linewidth]
  {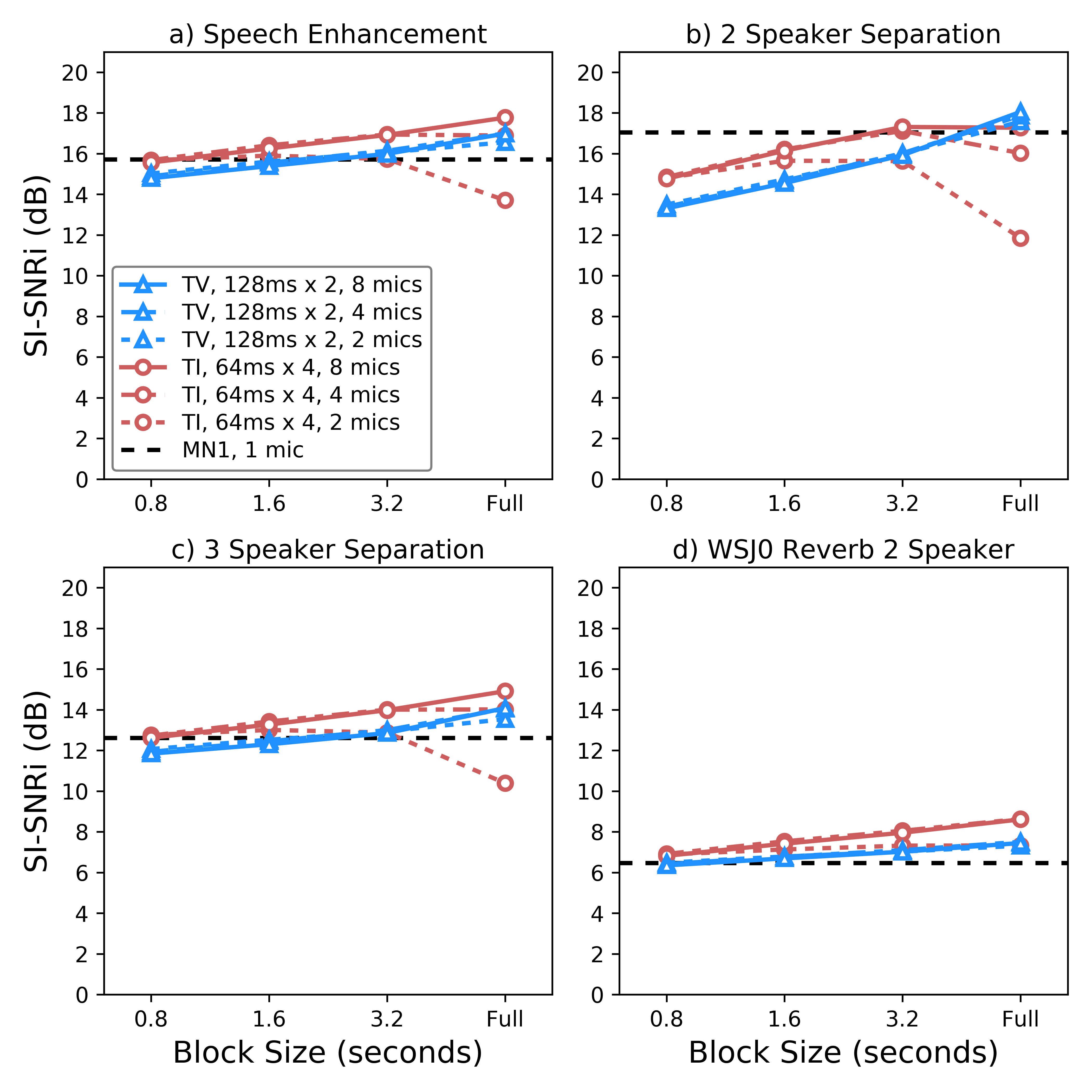}}
\vspace{-10pt}
\caption{Multi-frame beamforming {versus} block size.}
\label{fig:bf_multi}
\end{minipage}
\end{figure*}

\begin{figure*}[htb]
\centering
\begin{minipage}[b]{0.42\linewidth}
  \centering
  \centerline{\includegraphics[width=\linewidth]
{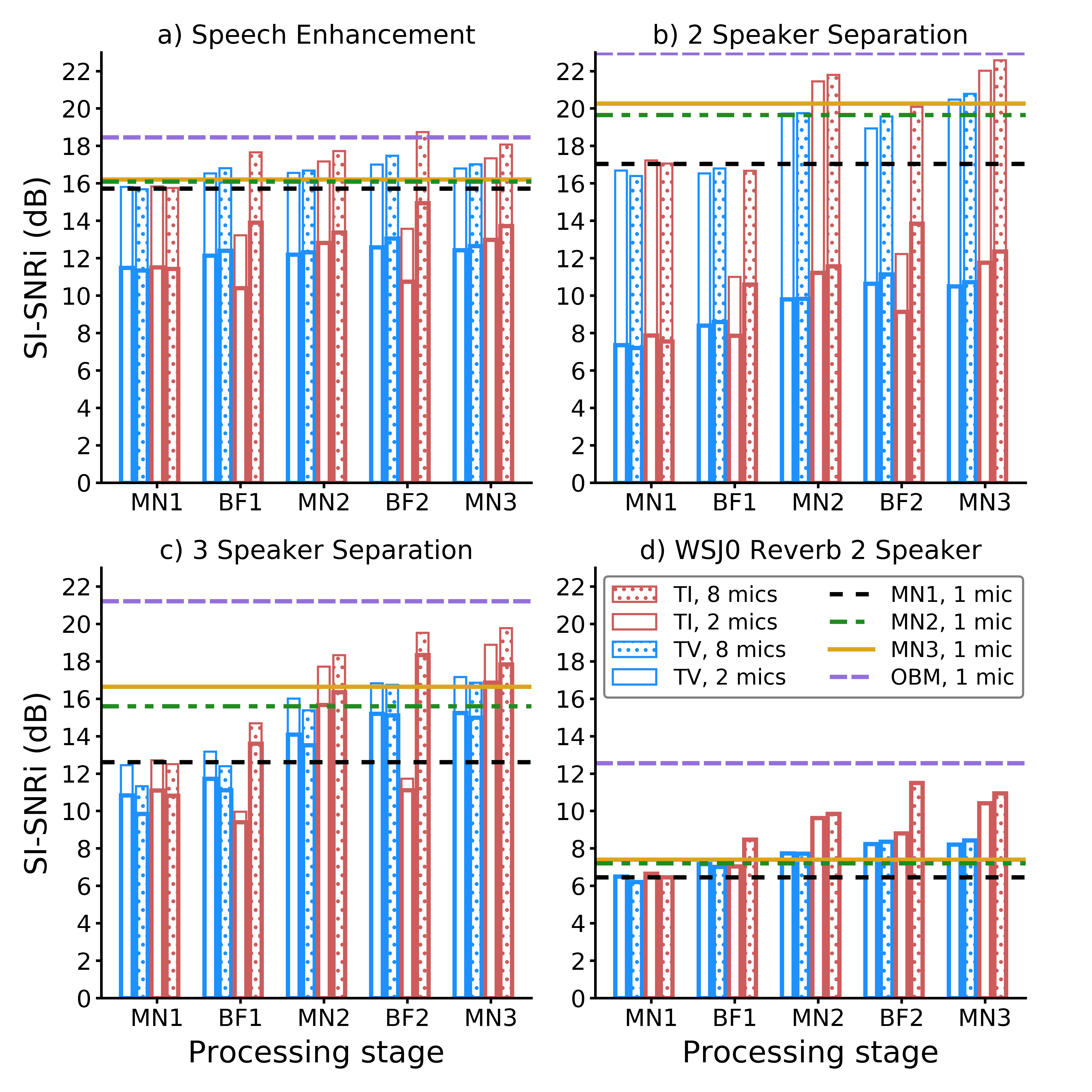}}
\vspace{-10pt}
\caption{
SI-SNRi of sequential neural BF (val). Dark\; bars indicate score of overlapping regions.
}
\label{fig:iter_val}
\end{minipage}
\hspace{0.025\linewidth}
\begin{minipage}[b]{0.42\linewidth}
  \centering
  \centerline{\includegraphics[width=\linewidth]
  {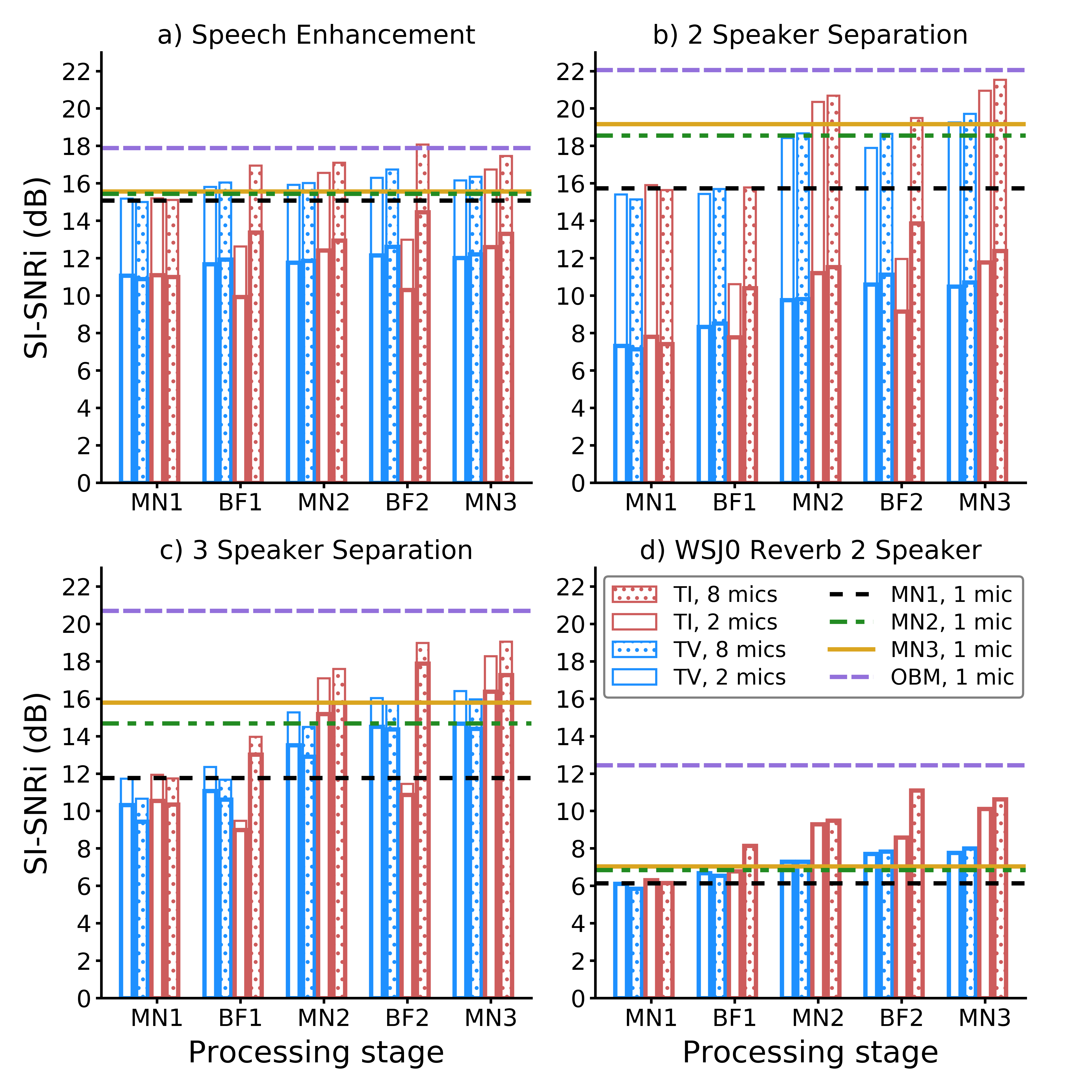}}
\vspace{-10pt}
\caption{SI-SNRi of sequential neural BF (test). Dark\; bars indicate score of overlapping regions.}
\label{fig:iter_test}
\end{minipage}
\end{figure*}

\begin{figure*}[htb]
\centering
\begin{minipage}[b]{0.42\linewidth}
  \centering
  \centerline{\includegraphics[width=1.04\linewidth]
{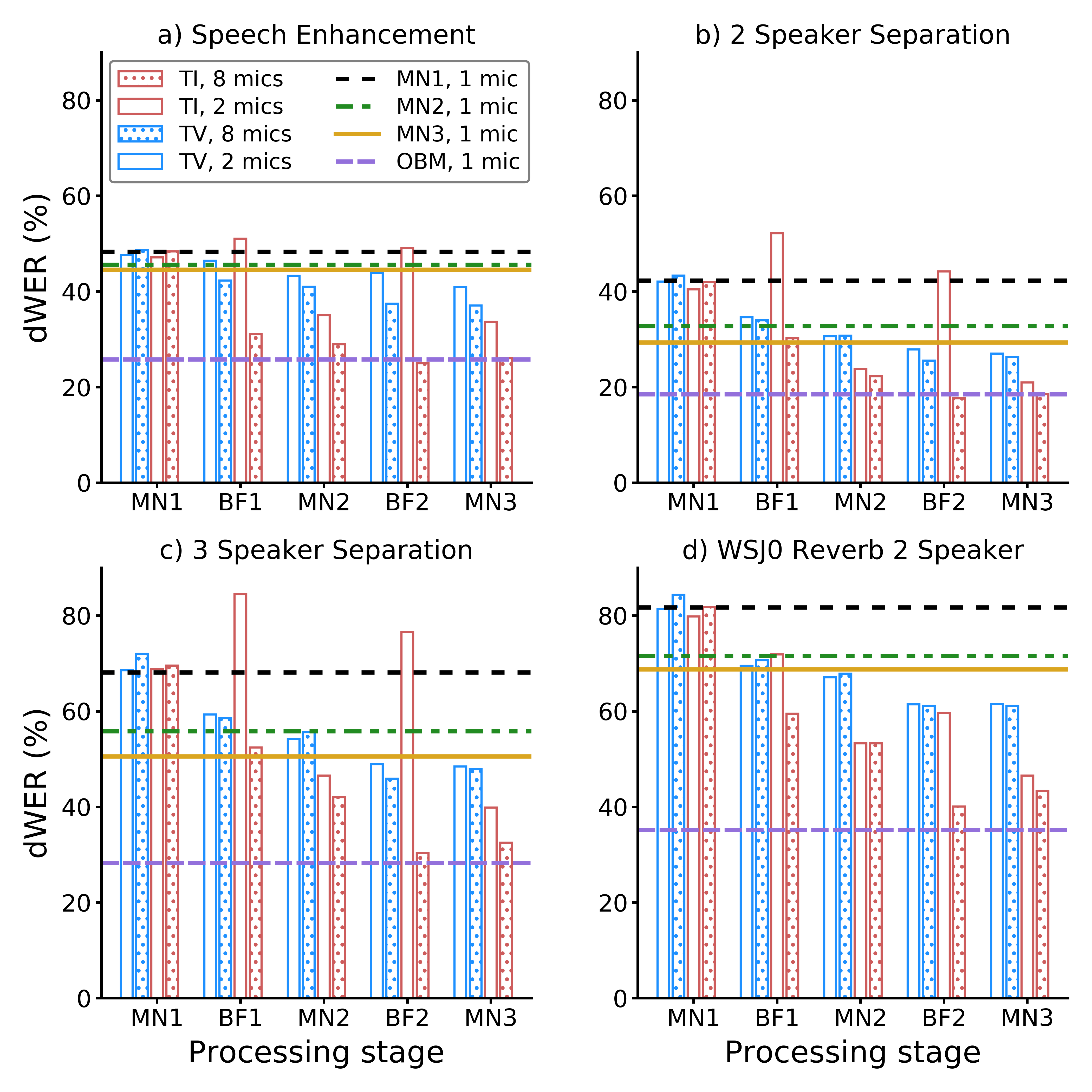}}
\vspace{-10pt}
\caption{Differential WER versus iteration (val).}
\label{fig:dwer_val}
\end{minipage}
\hspace{0.025\linewidth}
\begin{minipage}[b]{0.42\linewidth}
  \centering
  \centerline{\includegraphics[width=1.04\linewidth]
{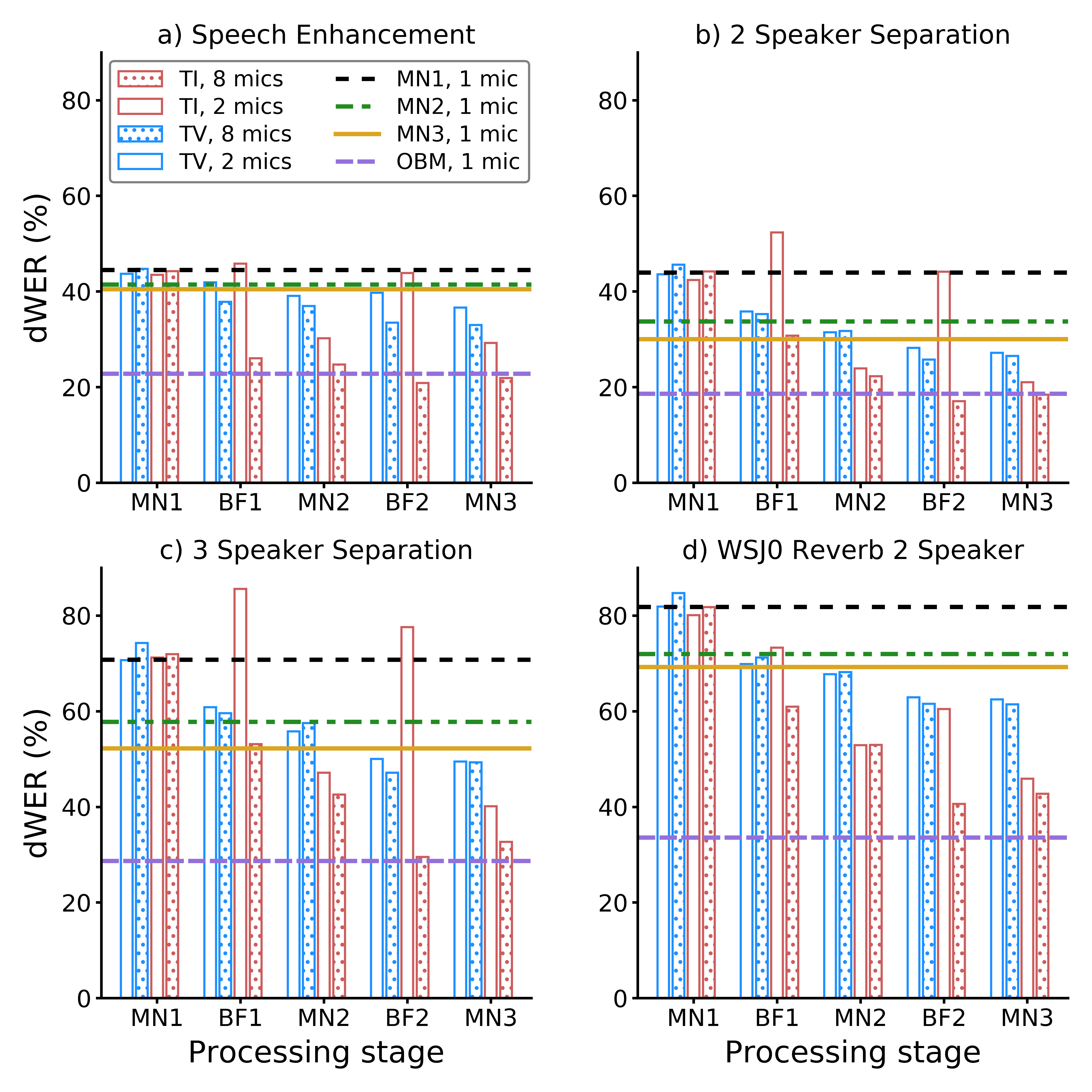}}
\vspace{-10pt}
\caption{
Differential WER versus iteration (test).
}
\label{fig:dwer_test}
\end{minipage}
%
\end{figure*}

\begin{table*}[h!]
    \centering
    \scalebox{0.8}{
    \begin{tabular}{l|cc|cc|cc|cc|cc|cc|cc|cc}
    \hline\hline
    & \multicolumn{8}{c|}{SI-SNRi (dB)} & \multicolumn{8}{c}{dWER (\%)}\\
    Method &   \multicolumn{2}{c|}{Speech} & \multicolumn{2}{c|}{2 Speaker} &   \multicolumn{2}{c|}{3 Speaker} &  \multicolumn{2}{c|}{WSJ0 2 Spk. } &  \multicolumn{2}{c|}{Speech} & \multicolumn{2}{c|}{2 Speaker} & \multicolumn{2}{c|}{3 Speaker} & \multicolumn{2}{c}{WSJ0 2 Spk. } \\
    & \multicolumn{2}{c|}{Enhancement} & \multicolumn{2}{c|}{Separation} &  \multicolumn{2}{c|}{Separation}  & \multicolumn{2}{c|}{Separation}   & \multicolumn{2}{c|}{Enhancement} & \multicolumn{2}{c|}{Separation} &  \multicolumn{2}{c|}{Separation}  & \multicolumn{2}{c}{Separation}   \\
     & val & tst & val & tst & val & tst & val & tst & val & tst & val & tst & val & tst & val & tst \\
    \hline
    Noisy & -  & -  & -  & -  & -  & -  & -  & -  & 71.5 & 67.1 & 99.8 & 98.6 & 144.6 & 143.4 & 111.1 & 111.7 \\
    \hline
    MN3, 1 mic & 16.2 & 15.6 & 20.3 & 19.2 & 16.7 & 15.8 & 7.4 & 7.0 & 44.6 & 40.5 & 29.3 & 30.1 & 50.6 & 52.3 & 68.8 & 69.3 \\
    BF1, 8 mic, TI 128ms x 1 & 15.7 & 15.2 & 16.1 & 15.5 & 14.4 & 13.7 & 8.6 & 8.3 & 31.1 & 26.6 & 27.8 & 27.5  & 52.0 & 52.6  & 57.1 & 58.2 \\
    \hline
    BF2, 8 mic, TI 64ms x 4  & \bf{18.7}* & \bf{18.1}* & 20.1 & 19.5 & 19.5 & \bf{19.0} & \bf{11.5} & \bf{11.1} & \bf{25.0}* & \bf{20.9}* & \bf{17.6}* & \bf{17.0}* & \bf{30.3} & \bf{29.5} & \bf{40.1} & \bf{40.6} \\
    MN3, 8 mic, TI 64ms x 4 & 18.1 & 17.5 & \bf{22.6} & \bf{21.5} & \bf{19.8} & \bf{19.0} & 10.9 & 10.6 & 26.1 & 21.9 & 18.5 & 18.4 & 32.5 & 32.7 & 43.3 & 42.8 \\
    \hline
    OBM, 1 mic, {\bf{oracle}} mask & 18.5 & 17.9 & 23.0 & 22.1 & 21.2 & 20.7 & 12.6 & 12.4 & 25.8 & 22.8 & 18.5 & 18.6 & 28.3 & 28.7 & 35.2 & 33.6  \\
    \hline\hline
    \end{tabular}
    }
    \caption{SI-SNRi and differential WER (dWER) results of sequential neural beamforming on the validation and test data as compared to strong single-channel and 8-mic multichannel baselines, as well as an oracle binary mask (OBM) for four different tasks. $^*$ indicates performance better than the oracle binary mask.}
    \label{tab:results}
    \vspace{-20pt}
\end{table*}

\subsection{Networks}
The network architecture of the TDCN++ networks 
is similar to the recently proposed Conv-TasNet \cite{Luo2019} and includes a few improvements introduced in \cite{kavalerov2019universal}. It consists of 4 repeats of 8 layers of convolutional blocks. Each block consists of a dilated separable convolution with feature-wise 
global layer normalization and a residual connection, where the dilation factor for the $k$th block is $2^k$. In contrast to Conv-TasNet, we utilize STFT basis with 32 ms windows rather than a learned basis with a very small window size, as initial results showed that the former leads to better performance. This is likely because an STFT with a larger window can better deal with room reverberation. The hop size is 8 ms. The sampling rate is 16 kHz. A 512-point FFT is used to extract 257-dimensional magnitude features for mask estimation. We use $\tau=10^{-3}$ and $\epsilon=10^{-8}$ with the soft-thresholded and stabilized negative SNR loss in (\ref{sc-loss}). We intend to open-source our implementations of TDCN++ and sequential multichannel models. 
SI-SNR
improvement (SI-SNRi) \cite{LeRoux2018a} over unprocessed speech is utilized as the evaluation metric. 
We also report {\em differential} word error rates (dWER) by assuming the speech recognition hypothesis of clean speech as the ground truth, and calculating the WER between the recognition output of enhanced/separated speech and that of clean speech.
The ASR model we used for obtaining differential WERs on the simulated datasets is an attention-based encoder-decoder model with 16k word-piece output units trained on 960 hours of Librispeech data \cite{irie2019unit}. For LibriCSS experiments, we used a kaldi based hybrid-HMM ASR model and an end-to-end ASR model based on ESPNet.

As a single-channel baseline, we consider a sequential TDCN++ network \cite{kavalerov2019universal}, where no spatial information is used. This network consists of three masking networks. For the second and third networks, the separated time-domain outputs of the previous network are concatenated with the time-domain mixture signal as the input features 
to produce separated estimates. We report performance for the output of each stage. This model is trained with the negative stabilized SNR loss in (\ref{sc-loss}) on the separated waveforms of all three stages.

\section{Results}
\label{sec:results}





\subsection{Results on simulated data}
Figures \ref{fig:bf_single} and \ref{fig:bf_multi} show the performance on the validation set of beamforming methods driven by a single-channel neural network under different conditions: using either TI or TV covariance estimation, for various block sizes, with 2, 4, or 8 microphones, for each of the four tasks, and with single (Figure \ref{fig:bf_single}) or multiple context frames (Figure \ref{fig:bf_multi}). We only display results for the best beamforming parameters over all tasks and numbers of microphones. We considered beamforming window sizes of 32, 64, 128 and 256 ms with half-sized hops. Frame context size were swept in powers of 2, and we chose the most promising frame context for each window. The best multi-frame TI result is achieved with window size 64 ms and 4 context frames (64 x 4), and the best multi-frame TV result uses window size 128 ms and 2 context frames (128 x 2). Our frame contexts are centered around the current frame, where for even context sizes, left context $a$ is one larger than the right context $b$.

In the following discussion, we use MN$i$ to refer to the mask network output after step $i$ and BF$i$ to refer to the beamformer applied after step $i$, as consistent with the subscripts in equations in Section \ref{sec:method}. For example, MN3 is the output of the separation model, and BF2 is the beamformed result achieved one step before that.

Figures \ref{fig:iter_val} and \ref{fig:iter_test} visualize the SI-SNRi performance on the validation and test sets of our best sequential neural beamforming models versus iteration, where the best multi-frame beamforming parameters are chosen from Figure \ref{fig:bf_multi}. These plots also display the performance of single-channel baselines, including a single-channel iterative network and an oracle binary mask (OBM) both using the same STFT parameters. Notice that performance generally improves monotonically with iterations, with the outputs of the neural networks achieving better SI-SNRi compared to the beamforming outputs. Also, despite performing worse than TV on their own, TI beamforming performs best when used in a sequential setup. For all tasks, using more than one microphone improves performance.

The speech enhancement and speech separation datasets that we constructed from Libri-Light \cite{kahn2020librilight} and \url{freesound.org} have less overlap between sources compared to the WSJ0 speech separation dataset. To fairly compare results, figures \ref{fig:iter_val} and \ref{fig:iter_test} shows SI-SNRi computed only on fully-overlapping segments with darker colored sub-bars. Notice that the results are more comparable between our speech separation and WSJ0 speech separation in overlapped regions. Also, for speech enhancement and 2 speaker separation, our sequential neural beamformer exceeds the performance of the oracle binary mask (not shown) in fully overlapped segments.

Figures \ref{fig:dwer_val} and \ref{fig:dwer_test} display dWER for the validation and test sets. Our sequential neural beamforming models significantly decrease dWER, especially when more microphones are used. When using TI beamforming with 8 microphones, BF2 achieves the best dWER as opposed to MN3 since ASR models tend to work better with linear time-invariant processed signals. For two microphones though, MN3 output is the best, likely because two-microphone beamforming cannot achieve sufficient spatial separation. For the Libri-Light+Freesound speech enhancement and speech separation tasks, the best-performing outputs of our model achieve comparable or slightly better dWER than an oracle binary mask.

Table \ref{tab:results} presents the results of our best eight-microphone system as compared to a single-channel baseline, a multichannel baseline, and an oracle binary mask. 
We point out that our baselines are strong ones since we use a state-of-the-art neural network architecture and an improved SNR loss function.
Also, for the beamformer, we use an optimal 128 ms window size which is typically not the case. Our methods obtain significantly better SI-SNR and dWER against these strong 
baselines.

\begin{table*}[h]
\centering
\caption{Performance of separation methods on LibriCSS eval set in terms of the resulting downstream diarization error rate (DER) (using spectral clustering), cpWER (using a hybrid HMM-DNN model) and cpWER (using an E2E model) results. Separation performance in terms of signal to distortion ratio (SDR) is reported on a different simulated LibriCSS-like eval set. For comparison, we also show results obtained on a ``no separation'' baseline.}
\label{tab:sep_result}
\begin{tabular}{@{}lcccc@{}}
\toprule
\textbf{Method} & \textbf{SDR (dB)} & \textbf{DER (\%)} & \textbf{HMM-DNN cpWER (\%)} & \textbf{E2E cpWER (\%)}\\ \midrule
No separation & - & 18.28 & 31.04 & 27.11 \\
Mask-based MVDR \cite{chen2020continuous} & 5.8 & \textbf{13.86} & 22.75 & 13.37  \\
Proposed: sequential multi-frame (BF2) & \textbf{14.1} & 14.07 & \textbf{19.28} & \textbf{12.70} \\ \bottomrule
\end{tabular}
\vspace{-1em}
\end{table*}

\subsection{Results on LibriCSS}
We ran our three-speaker separation model on the LibriCSS evaluation dataset which contains 54 different 10 minute meetings with varying amounts of overlap recorded with a circular 7-microphone array \cite{chen2020continuous}. We evaluate our separation model within a separation-diarization-recognition pipeline which is described in more detail in \cite{raj2020}. The model is applied in 8 second overlapping blocks with a 4 second shift. 7 microphone data is padded with another channel obtained by shifting the first microphone signal by one sample and adding white Gaussian noise with variance 1e-6. We used the BF2 output of the three-speaker separation model since it achieved better dWER in our experiments on simulated test data as can be seen in Table \ref{tab:results}. After separating each block into three speaker tracks, we stitch together the separated block-length tracks into meeting-length tracks. We use magnitude STFT domain mean-squared distance between common parts of neighboring blocks to find the best permutation between them. 
Diarization was done using x-vector based segment clustering from multiple separated tracks with some post-refinement \cite{raj2020}.
Each diarized segment is recognized using either a hybrid HMM-DNN ASR model or an end-to-end (E2E) ASR model. 

The results, taken from \cite{raj2020} are shown in Table \ref{tab:sep_result}. We compare with a baseline mask-based MVDR separation method that uses bidirectional LSTM layers \cite{chen2020continuous, raj2020}. 
Separation performance was evaluated on 
a simulated test set 
since reference signals are required \cite{raj2020}. 
Separated tracks are first mapped to $N$-speaker tracks where $N$ is the number of participating speakers in a meeting, using a method we call ``oracle track mapping'' \cite{raj2020}. 
Average meeting-level SDR~\cite{vincent2006performance} is reported.
Multi-frame MCWF beamformer (BF2) achieved a much better SDR value as compared to mask-based MVDR, however this may be expected since the MVDR beamformer does not attempt to reconstruct the target signal at the reference microphone directly. 

On LibriCSS, diarization error rate (DER) was close between two separation methods. Our model achieved a concatenated minimum-permutation WER (cpWER)~\cite{Watanabe2020CHiME6CT} of 19.28\% on the LibriCSS eval set using a hybrid HMM-DNN model better than the baseline system. When using a superior E2E ASR model, our separation model achieved a cpWER of 12.70\% as compared to a baseline of 13.37\%. Note that our model is trained using a completely different microphone geometry and with mismatched data in terms of overlap amount. However, our model has the advantage of training from on-the-fly mixtures and Libri-Light database which is quite large. We can see that the model generalizes to unseen conditions since it works well on data which contains a different microphone geometry, unseen source locations, different overlap amounts, and unseen reverberations with real recording conditions.

\section{Conclusions}
\label{sec:conclusions}
We have explored an alternating strategy between spectral estimation using a mask-based network and spatial estimation using beamformers. For spatial estimation, we introduced multi-frame beamforming and compared multiple ways of computing covariance matrices for time-invariant and time-varying beamforming.
Evaluation results on four sound separation tasks suggest that, when combined with neural network based mask estimation, time-invariant multi-frame beamforming with a reasonably large window and context size produces the best separation performance for non-moving sources. 
Our best three-stage method demonstrates an average improvement of 2.75 dB in SI-SNR and an absolute reduction of 14.2\% in dWER over several strong and representative baselines, across four challenging reverberant speech enhancement and separation tasks. Our three speaker separation model was used to separate tracks from LibriCSS evaluation dataset and ended up improving the constrained permutation word error rate as compared to a mask-based MVDR beamformer baseline.

\textbf{Acknowledgment.} LibriCSS part of this work was done at JSALT 2020 at JHU, with support from Microsoft, Amazon, and Google.
\newpage
\newpage
\bibliographystyle{IEEEtran_nourl}
\balance
\bibliography{refs}

\end{document}